\documentstyle[11pt]{article}
\def\thefootnote{\fnsymbol{footnote}}
\begin{document}
\newcommand{\eqn}[1]{(\ref{eq:#1})}
\newcommand{\eq}{\begin{equation}}
\newcommand{\en}{\end{equation}}
\newcommand{\bea}{\begin{eqnarray}}
\newcommand{\eea}{\end{eqnarray}}
\newcommand{\nn}{\nonumber \\ }
\newcommand{\bdm}{\begin{displaymath}}
\newcommand{\edm}{\end{displaymath}}
\newcommand{\A}{\cal{A}_\gamma }
\newcommand{\At}{\tilde{\cal{A}}_{\gamma}}
\newcommand{\Ak}{{\cal A}_{3k}}
\newcommand{\SU}{\widehat{SU(2)}}
\newcommand{\SP}{\widehat{SU(2)^+}}
\newcommand{\SM}{\widehat{SU(2)^-}}
\newcommand{\ba}{\begin{array}}
\newcommand{\ea}{\end{array}}
\newcommand{\ds}{\displaystyle}
\newcommand{\ZZ}{\hbox{{\rm Z{\hbox to 3pt{\hss\rm Z}}}}}
\newcommand{\SSS}{\widehat{SU(3)}_{\tilde{k}^+}}
\newcommand{\br}{\langle}
\newcommand{\kt}{\rangle}
\newcommand{\um}{\frac12}
\newcommand{\bra}[1]{\langle {#1}|}
\newcommand{\ket}[1]{|{#1}\rangle}
\newcommand{\lm}{\ell^-}
\newcommand{\lp}{\ell^+}
\newcommand{\la}{\lambda}
\newcommand{\al}{\alpha}
\newcommand{\eps}{\epsilon}
\newcommand{\vl}{\vec{\lambda}}
\newcommand{\pa}{\partial}
\newcommand{\ktp}{\tilde{k}^+}
\newcommand{\ktm}{\tilde{k}^-}
\newcommand{\sq}{{\sqcup}}
\newcommand{\cd}{{\cal D}}
\def\spinst#1#2{{#1\brack#2}}
\def\CG{{\cal G}}
\def\t{$\times$}
\newcommand{\NP}[1]{Nucl.\ Phys.\ {\bf #1}}
\newcommand{\Prp}[1]{Phys.\ Rep.\ {\bf #1}}
\newcommand{\PL}[1]{Phys.\ Lett.\ {\bf #1}}
\newcommand{\NC}[1]{Nuovo Cim.\ {\bf #1}}
\newcommand{\CMP}[1]{Comm.\ Math.\ Phys.\ {\bf #1}}
\newcommand{\PR}[1]{Phys.\ Rev.\ {\bf #1}}
\newcommand{\PRL}[1]{Phys.\ Rev.\ Lett.\ {\bf #1}}
\newcommand{\MPL}[1]{Mod.\ Phys.\ Lett.\ {\bf #1}}
\newcommand{\IJMP}[1]{Int.\ J.\ Mod.\ Phys.\ {\bf #1}}
\newcommand{\JETP}[1]{Sov.\ Phys.\ JETP {\bf #1}}
\newcommand{\TMP}[1]{Teor.\ Mat.\ Fiz.\ {\bf #1}}

\begin{titlepage}
%\null
\begin{flushright}
DFTT--46/95\\
{\tt hep-lat/9510019}
\end{flushright}
%\vspace{1cm}
\vskip1.0cm
\begin{center}
{\Large\bf Width of Long Colour Flux Tubes\\
\vskip .5 cm
in Lattice Gauge Systems}
\end{center}
\vskip 0.8cm
\centerline{M. Caselle$^a$, F. Gliozzi$^a$, U. Magnea$^{a,b}$ and 
S. Vinti$^{a}$}
\vskip 0.8cm
\centerline{\sl  $^a$ Dipartimento di Fisica
Teorica dell'Universit\`a di Torino}
\centerline{\sl Istituto Nazionale di Fisica Nucleare, Sezione di Torino}
\centerline{\sl Via P.Giuria 1, I--10125 Torino, Italy
\footnote{e--mail: caselle, gliozzi, vinti~@to.infn.it}}
\vskip 0.6cm
\centerline{\sl $^b$State University of N.Y. at Stony Brook, Stony Brook,
NY 11794 USA
\footnote{e--mail: blom~@to.infn.it}}
\vskip 1. cm

\begin{abstract}
\vskip .6 cm
In the confining phase of any gauge system the mean squared width 
of the colour flux tube joining a pair of quarks
should grow logarithmically as a function of their distance, 
according to the effective string description of its infrared 
properties. 
New data on $3D$ $\ZZ_2$ gauge theory, combined with high precision 
data on the interface physics of the $3D$ Ising model fit nicely this 
behaviour over a range of more than two orders of magnitude.
\end{abstract}
\vskip 1.5em
\end{titlepage}
\baselineskip=0.8cm
\renewcommand{\thefootnote}{\arabic{footnote}}
\setcounter{footnote}{0}

\section{Introduction}
\vskip .3 cm

The mechanism of confinement for abelian and non--abelian gauge 
theories in three or four space--time dimensions seems 
well described by the old conjecture \cite{tmp} that the vacuum 
behaves as a dual superconductor.
The dual Meissner effect squeezes the colour field generated by a pair 
of quark sources  into a thin flux tube (the dual version 
of the Abrikosov vortex), generating a static potential proportional to 
the interquark distance. Numerical studies of the distribution of the 
colour flux around the quark sources \cite{ft,dg,bss} seem to support 
this picture.

According to another old conjecture \cite{no,lsw,lmw}, this flux tube 
can vibrate as a free string. There are many interesting and impressive 
quantum effects associated to these string vibrations.
One of them seems to conflict with a property of the 
dual Abrikosov vortices:  the intrinsic thickness of these is a 
physical quantity which is independent of the interquark distance and 
is fixed by the parameters of the Ginzburg--Landau formulation of the 
superconductivity, while in the string picture the width of the flux 
tube grows logarithmically as a function of the interquark 
distance \cite{lmw}. 

In this paper we will show compelling numerical evidence of this 
logarithmic growth. However this does not mean that the picture of the dual 
Abrikosov vortex does not work. Simply, the way one defines the mean 
squared width of the flux tube in the lattice simulation  also takes  
into account the quantum effect of the vibration modes of the dual 
Abrikosov vortex, yielding an effective width which is larger than the 
intrinsic thickness of a non--vibrating vortex.

It is worthwhile to stress that the dual superconductor nature of the 
confining vacuum and the string--like behaviour of the colour flux tube 
are conceptually two independent properties of the theory. Indeed in 
the strong coupling regime, before the roughening transition, {\it i.e.}
in the region where the strong coupling expansion of the string tension 
$\sigma$ converges, the dual Meissner effect is already operating and 
the colour flux tube connecting  pairs of quark sources has a 
string-like form, however cannot vibrate. Hence in this strong coupling 
phase  the dual Abrikosov vortices, like those of the ordinary 
superconducting media, do not exhibit any kind of quantum string effects.

At the roughening point, the colour flux tube of whatever $3D$ or $4D$ 
gauge theory undergoes a transition towards a rough phase (which is the 
one  connected to the continuum limit), where it can 
vibrate freely and the energy stored in its vibration modes is no longer 
negligible. The collective degrees of freedom describing such a 
string--like motion can be described by a $2D$ field theory known as 
{\it effective string theory}.

It is widely believed that the roughening transition is described by the 
universality class of Kosterlitz and Thouless \cite{kt}. Accordingly,
the renormalization group equations show that in the rough phase the 
effective string theory 
describing the dynamics of the flux tube flows  at large 
scales towards a massless free field theory. Thus, for large enough 
interquark separations,  it is not necessary to know explicitly 
the specific form of the effective string action describing 
the behaviour of the  flux tube,   but only its infrared massless
limit. This simple, general fact has some important implications because 
the limit action is  critical in the whole rough phase.  
As a consequence, it produces universal ({\it i.e.} gauge group 
independent) finite size effects. Moreover, because this theory is also 
infrared divergent, such finite size effects are expected to be rather 
strong. 

The effects we shall describe require highly accurate 
numerical data that at present can hardly be reached in 4D $SU(2)$ 
and $SU(3)$ gauge models.

Luckily, these effects can be checked accurately in the $3D$ $\ZZ_2$ 
gauge model which is the simplest prototype of gauge theory. 
\vskip .3 cm
\section{ Width of the Colour Flux Tube }
The logarithmic behaviour of the mean squared width $w^2(R)$ of the 
colour flux tube as a function of  the interquark distance $R$ has been 
predicted many years ago by  L\"uscher, M\"unster and Weisz \cite{lmw} 
in the framework of the effective string picture of  gauge systems. We 
shall rederive and refine  this universal law by  directly using some 
exact results on the two dimensional free gaussian model in a finite box 
and compare it with Monte Carlo simulations on the $3D$ $\ZZ_2$ 
gauge model.

We shall see that the free gaussian model describes quite accurately 
this phenomenon.

\subsection{{\sl Results from the Gaussian Limit }}

In order to fix  notations, let us recall that the simplest way to 
implement the effective string picture of the colour flux tube is
to assume that the vacuum expectation value of a large Wilson loop 
$W(C)$ can be represented by the functional integral
\eq
\br W(C)\kt=\int\prod_{i=1}^{D-2}[Dh_i]~{\mathrm e}^{-
\int{\mathrm d}^2\xi\,{\cal L}(h_i)}~~.
\en
For simplicity $C$ is taken  planar  and $\xi^\alpha,\,\alpha=1,2$ 
are the coordinates of the planar domain $\cd$ bounded by $C$. 
$h_i(\xi^1,\xi^2)$ are the transverse coordinates of the string 
describing the configurations of the flux tube in its space--time 
evolution. They are the fields entering in the effective string action 
$S={\int{\mathrm d}^2\xi\,{\cal L}(h_i)}$, which is largely unknown, 
but is believed to flow to the free gaussian 
model for very large loops $C$, 
as explained in the introduction. More precisely we have
\eq
S\to\sigma A+\int_\cd{\mathrm d}^2\xi\,\sum_{i=1}^{D-2}
\frac{\sigma}2
~(\partial_ih_i)^2~~~,
\label{gali}
\en
where $\sigma$ is the string tension and 
$A=\int_\cd d^2\xi$ is the area of the planar domain.  
Eq.~(\ref{gali}) can be considered as the first two terms of an expansion 
in the adimensional parameter $1/\sigma A$. Recently, the first 
non--gaussian correction of such an expansion has also been 
studied \cite{pv,inter2}.

The mean squared  width of the  flux tube 
generated by this Wilson loop  is defined as the sum of 
the mean square deviations of the transverse coordinates 
$h_i(\xi^1,\xi^2)$ of the underlying string, {\it i.e.}
\eq
w^2(\xi^1,\xi^2)=\sum_{i=1}^{D-2}
~\br\left(h_i(\xi^1,\xi^2)-h_i^{CM}\right)^2\kt~~~,
\label{w2}
\en
where $h_i^{CM}$ is the transverse coordinate of 
the center of mass of the flux tube. In terms of the Green functions 
$G_i$ we get
\eq
\sigma w^2(\xi)=\sum_{i=1}^{D-2}\left\{G_i(\xi,\xi+\varepsilon)-
\frac2{A}\left[\int_\cd d^2\xi'\left(G_i(\xi,\xi')-
\frac1{2A}\int_\cd d^2\xi"G_i(\xi",\xi')\right)\right]\right\}~~~,
\label{w22}
\en
where $\varepsilon$ is a UV cut--off, $\sigma$ is the string tension and 
the Green function is defined as
\eq
G_i(\xi,\xi')=\sigma~\br h_i(\xi)\,h_i(\xi')\kt~~~.
\en
The vacuum expectation value is taken with respect to 
the string action $S$. In the infrared limit the Green function 
fulfills the free field equation
\eq
-\Delta G(\xi,\xi')=\delta^{(2)}(\xi-\xi')~,
\en
where for simplicity the index $i$ of the transverse direction has been 
omitted. 
Instead of considering the usual expansions of  $G$ in terms of the 
eigenfunctions of $\Delta$ it is possible (and much more useful) 
to write it in a more compact form. 

Indeed the problem of finding the  Green function for an arbitrary, 
simply connected region $\cd$  can be solved in a closed form once 
a conformal mapping $z=\xi^1+i\,\xi^2\to \ell$ of $\cd$ onto 
the unit circle $\vert\ell\vert=1$ is found which maps $z'=
\xi^{\prime1}+i\,\xi^{\prime2}$ 
into the origin  $\ell=0$. Denoting by $\ell_{z^\prime}(z)$ 
the analytic function providing us with such a  mapping, it is 
immediate to verify that the real function 
$f_{z'}(z,\cd)=\log\vert\ell_{z'}(z)\vert$ is harmonic in the 
punctured set $\cd\setminus\{z'\}$~ (hence $\Delta f=0$ for 
$\;z\not=z'$), 
vanishes at the boundary $\partial\cd$ and diverges logarithmically as 
$\log\vert z-z'\vert$ for $z\to z'$. It follows that the Green 
function is given by
\eq
G_{\cd}(z,z')=-\frac1{2\pi}f_{z'}(z,\cd)~~~.
\en

Some explicit examples can be found in Tab.~I and will be discussed 
later.

Inserting such expressions in Eq.(\ref{w22}), we can evaluate analytically 
the shape of the flux tube as a function of the coordinate $z$ inside the
domain $\cd$. One can observe a very steep increase of $w^2(z)$ at the 
boundary of $\cd$ followed by a plateau covering approximately the $90\%$ of 
the domain, where
$w^2(z)$ grows very slowly an reaches its maximum in the center $z_o$ of the
domain. For instance, in the case of an infinite strip of width $R$ we get
\eq
2\pi\sigma w^2(r)=\log\frac{\vert\sin(2r+\frac{R_c}{R})\vert}
{\vert\sin(\frac{R_c}{R})\vert}~~~,
\en
where $r=\frac{\pi}{2R}x$,  $0\leq x\leq R$, and the UV cut-off has been 
reabsorbed in the length scale $R_c=\frac{\pi\varepsilon}2$.
 As a measure of the total width of the flux tube it is often used in the
literature the integrated quantity
\eq
w^2=\frac1A\int_\cd d^2\xi\; w^2(\xi^1,\xi^2)~~~.
\en
However in the applications to the gauge models this quantity may be affected by
non-universal corrections due to the self-interactions of the quark line at the
boundary. For this reason we prefer to use instead the value of the mean squared
width measured at the symmetry point of the domain:
\eq
w^2_o=w^2(\xi_o^1,\xi_o^2)~~~.
\label{w222}
\en
For instance, in the  Wilson loop of size $R\times T$ we have
obviously $\xi_o^1=\frac R2$ and $\xi_o^2=\frac T2$.
\vskip .2 cm
\begin{center}
%%%%%%%%%%%%%%%%%%%%%%%%%%%%%%%%%%%%%%%%%%%%%%%%%%%%%%%%%
\centerline{\sl Tab.~I}
\vskip .2 cm
\begin{tabular}{cccc}
\hline
$\cd$&b.c.&Green function&$2\pi\sigma\Delta w^2$\\
\hline
\\
Disk & {\it fixed}~~ &
$G_D(z,z')=-\frac1{2\pi}\log\left\vert R\frac{z-z'}
{R^2-z\bar{z}'}\right\vert$&$0.8105$\\
\\
Square& {\it fixed}~~ &
$G_{R}(z,z')=-\frac1{2\pi}\log\left\vert\frac{\sigma(z-z')\sigma(z+z')}
{\sigma(z-\bar z')\sigma(z+\bar z')}\right\vert$ &$-0.0139$\\
\\
Strip&  {\it fixed}~~ &
$G_S(z,z')=-\frac1{2\pi}\log\left\vert\frac{\sin(\pi(z-z')/2R)}
{\sin(\pi(z+\bar{z}')/2R)}\right\vert$&$0.8587$\\
\\
Torus& {\it periodic}~~&$\CG(z-z')$&$0$\\
\\
\hline
\end{tabular}
\end{center}
\vskip .1 cm
\begin{center}
{\it The closed form of the Green functions for some simple domains 
is given together with the difference between the mean squared width as 
defined in Eq.s~(\ref{w22},\ref{w222}) and that relative to a square torus.}
\end{center}
\vskip .2 cm

For our purposes, the relevant property of the conformal mapping $\ell$ 
is that one can perform an arbitrary  scale transformation 
$z\to\Lambda z$ without destroying the conformal character of $\ell$.
More precisely we can write
\eq
f_{z^\prime}(z,\cd)=f_{\Lambda z^\prime}(\Lambda z,\cd_\Lambda)~~~,
\label{conf}
\en
where $\cd_\Lambda$ denotes the scaled domain. A direct consequence is 
that it is always possible to fix the area of 
the scaled domain $\cd_\Lambda$ to an arbitrary value, say 1, without 
changing the Green function. It follows that the integration of the 
finite part $G(z,z^\prime)$ in Eq.~(\ref{w22}) cannot depend on the size 
of the domain $\cd$ but only on its shape. On the contrary the UV 
divergent part gives a contribution which grows logarithmically with 
the size of the domain. This can be simply understood as follows:
if  $R$ is a typical linear dimension of the domain $\cd$, the 
scaling property described above implies that the cut--off should appear 
always in the ratio ${\varepsilon}/R$. On the other hand, for 
$\varepsilon<<R$  the Green function behaves like 
$-\frac1{2\pi}\log(\varepsilon)$, 
yielding the logarithmic law 
\eq
w^2_o=\frac1{2\pi\sigma}\log\left(R/R_c\right)~~~,
\label{log}
\en
where the UV cut--off has been absorbed in the definition of the scale 
$R_c$. 

Note that the above equation is meaningful only for $R>R_c$. 
This inequality has a simple physical meaning: 
the dynamics of the flux tube is described by a truly free field theory 
only at large distances; the cut--off $\varepsilon$ sets up the 
ultraviolet scale $R_c$ below which either the free--field approximation 
breaks down or the internal degrees of freedom of the flux tube become 
important. Thus $R_c$ gives us a very rough estimate of the intrinsic 
thickness of the flux tube.

The absolute value of $R_c$ is not directly calculable, 
being proportional to the UV cut--off. We can however evaluate the ratios 
among the $R_c$'s of different domains making the very mild assumption 
that the UV cut--off $\varepsilon$ cannot depend on the size and the 
shape of the $\cd$ domain (which are on the contrary  infrared 
properties). For later convenience, we shall use as  reference scale 
$R_p$ associated to a squared box with {\it periodic}
boundary conditions ({\it i.e.} a torus). 
Also in this case the the Green function $\CG$ 
can be expressed in a closed form. More generally, for a torus of
sides $R\times T$ we have (see for instance Ref.~\cite{inter2}):
\eq
\CG(z)=-\frac{1}{2\pi}\ln\Bigl|\sigma(z)\Bigl|~
+~\frac{\pi E_2(it)}{12R^2}~\Re e(z^2)+\frac{1}{2RT}
\Bigl(\Im m\;z\Bigr)^2~~~,
\label{sol}
\en
where $\sigma(z)$ is the Weierstrass sigma function, defined through the
infinite product
\eq
\sigma(z)=z\prod_{\omega\not=0}\left(1-\frac{z}{\omega}\right)
{\rm  e}^{z/\omega+\um(z/\omega)^2}~~,~~\omega=mR+inT~~~.
\label{sigma}
\en
The set $\{\omega\}$ is known as the period lattice and 
$E_2(it)$ is the first Eisenstein series with $t=T/R$. 
For more details see also the Appendix.

Denoting with $w^2_p(R)=\frac1{2\pi\sigma}\log\left(R/R_p\right)$ the mean 
squared width of the flux tube generated by taking $\cd$ to be the 
square box of reference, it follows that the difference 
$\Delta w^2(R)=w^2_{o}(R)-w^2_p(R)$ does not depend on $R$ but is a 
calculable function of the shape of $\cd$.   

In Tab.~I one can find the closed 
form of the Green function and the corresponding $\Delta w^2$ of some 
simple domains like a disk of radius $R$, a square of side $R$, an 
infinite strip of width $R$ and the reference torus of size $R\times R$ 
($R=T$).
It is important to notice that the Weierstrass sigma function entering 
the definition of the Green function of the square of side $R$ 
corresponds to a torus of double size. 
For a rectangle of size $R\times T$ the Green function has the same 
formal expression as the square in terms of $\sigma(z)$, which now is 
defined on a torus of size $2R\times2T$. 

The detailed  calculation of $w^2$ for the torus (which is 
interesting in the physics of fluid interfaces) and for the 
rectangular Wilson loop can be found in the Appendix. The result is 
drawn in Fig.~1, where the behaviour of $\Delta w^2$ for a rectangle as 
a function of the aspect ratio $t=T/R$ (dashed line) is compared with 
that of  a rectangular torus with the same shape $t$ (continuous line) 
and with the value for the infinite strip (dotted line). 
Note that the almost linear 
increase of $\Delta w^2$ for the torus agrees with the fact that for 
large $t$ the theory becomes almost one--dimensional and describes the 
interface of a $2D$ Ising model, where the mean squared width is known 
to grow linearly with the size. 

This behaviour of $\Delta w^2$ calculated in the free gaussian model  
as a function of the shape and of the boundary conditions (periodic 
and fixed) can be easily checked in the $2D$ $XY$ model (in Fig.~1,
open circles and triangles respectively) 
which is believed to belong to the Kosterlitz--Thouless universality 
class and certainly flows to the free gaussian model at large 
scales. The other numerical data reported in Fig.~1 (black dots) 
are data extracted from the $3D$ $\ZZ_2$ gauge model simulations  
which will be discussed in the next section.

If one is able to evaluate, besides the effective 
squared width, also the higher moments $w^{(2n)}$, defined by
\eq
w^{(2n)}=\frac1{A}\sum_{i=1}^{D-2}\int_\cd d^2\xi~
\br\left(h_i(\xi^1,\xi^2)-h_i^{CM}\right)^{2n}\kt~~~,
\label{w2n}
\en
one can retrieve the three-dimensional distribution of the colour field 
strength density $\rho(x,y,z)$ inside the flux tube.
Indeed, choosing for simplicity   the domain $\cd$  
sitting in the plane $z=0$, we can implicitly define such a density through 
the following equations
\eq
w^{(2n)}=\frac{\int \rho\, z^{2n} dx\,dy\,dz\,}{\int\rho\, 
dx\,dy\,dz\,}~~~.
\en
As a consequence, the Fourier transform of $\rho$ can be written in 
terms of these moments as
\eq
\tilde\rho(p)=\sum_{n=0}^{\infty}(-1)^n\frac{p^{2n}}{(2n)!}w^{(2n)}
~~,
\label{rf}
\en 
where the integral of the density has been normalized to 1.
In the free gaussian model the above calculation can be performed 
explicitly using the Wick theorem, yielding simply
\eq
w^{(2n)}=(2n-1)!!~(w^2)^n~~~.
\en
Inserting this result into Eq.~(\ref{rf}) one finds that the density 
has a gaussian distribution
\eq
\rho_g(z)=a\exp(-z^2/b^2)~~~.
\label{gaussian}
\en

\subsection{ {\sl Simulations on $\ZZ_2$ Gauge Model}}

Though the logarithmic growth of the squared width of the flux tube is 
the most important  and model--independent quantum effect predicted by 
the effective string description, it has not been observed until 
now, because it is an infrared phenomenon that can be seen
only in very large Wilson loops, which are at the limit of the 
sizes reached by  present numerical simulations on $SU(2)$ and 
$SU(3)$ $4D$ gauge theories.
However, it is nowadays possible to overcome this problem in the 
$\ZZ_2$ $3D$ gauge model by exploiting some special features of this 
model.

The $3D$ $\ZZ_2$ gauge 
model on a cubic lattice is defined through the partition function
\eq
Z_{gauge}(\beta)=\sum_{\{\sigma_l=\pm1\}}\exp\left(-\beta S\right)~~~,
\en
where the action $S$ is a sum over all the plaquettes of the cubic lattice
\eq
S=-\sum_{\Box}\sigma_\Box~~~,~~~
\sigma_\Box=\sigma_{l_1}\sigma_{l_2}\sigma_{l_3}\sigma_{l_4}~~~.
\en
This model can be translated into the usual $3D$ Ising 
model  by the usual Kramers--Wannier duality transformation 
\bea
Z_{gauge}(\beta)~\propto~ Z_{spin}(\tilde\beta)&&\\
\tilde{\beta}=-\um\log\left[\tanh(\beta)\right]~~&&~~,
\eea
where $Z_{spin}$ is the partition function of the Ising model on the 
dual lattice
\eq
Z_{spin}({\tilde\beta})=\sum_{s_i=\pm1}\exp(-\tilde\beta H(s))~~~,
\en
with
\eq
H(s)=-\sum_{\br ij \kt}J_{\br ij \kt}s_is_j
\en
where $i$ and $j$ denote nodes of the dual lattice and the sum is 
extended to the links ${\br ij \kt}$ connecting  the nearest--neighbour 
sites. For the moment the couplings $J_{\br ij \kt}$ are all chosen 
equal to $+1$ .

Using the duality transformation it is possible to build 
up a one--to--one mapping of physical observables of the gauge system 
onto 
the corresponding spin quantities. For instance, the vacuum 
expectation value of a Wilson loop $W(C)$ can be expressed in terms of 
spin variables as follows. First, choose an arbitrary surface $\Sigma$ 
bounded  by $C$: $\partial\Sigma=C$; then "frustrate" the links 
intersecting $\Sigma$, {\it i.e.} take $J_{\br ij \kt}=-1$ whenever 
$\br ij\kt\cap\Sigma\not=\emptyset$.
Let us denote with $H'(s)$ the 
Ising Hamiltonian with this choice of couplings: the new Ising partition 
function $Z_{spin}'({\tilde\beta})=
\sum_{s_i=\pm1}\exp\left(-\tilde\beta H'(s)\right)$
describes a vacuum modified by the  Wilson loop $W(C)$, which we shall
call the W--vacuum. A well--known consequence of duality is that 
\eq
\br W(C=\partial \Sigma)\kt_{gauge}
~=~{Z_{spin}'\over Z_{spin}}~=~
\br\prod_{\br ij\kt\cap\Sigma\not=\emptyset}
\exp(-2{\tilde\beta}s_is_j)\kt_{spin}~~~,
\label{wil}
\en
where  the product is over all the dual links intersecting $\Sigma$.

One of the main difficulties in dealing with numerical simulations is 
that near a critical point the successive configurations  generated by a 
Monte Carlo  procedure are strongly correlated. Then it is hard 
to get a reliable  set of thermalized configurations where the 
measurements  
can be  done with sufficient accuracy. This phenomenon is known as 
critical slowing down.

A great advantage of mapping a gauge observable onto a spin observable 
is that a  non--local cluster 
updating algorithm \cite{sw} can be used  in place of the 
usual Metropolis or heat--bath methods. 
This kind of algorithm substantially reduces critical slowing down. 

This allows us to probe the structure of the flux tube with 
a high accuracy. There is no analogous procedure 
for the other gauge theories. 

The procedure is the following. A Wilson loop 
$W(C)$ is realized in the spin lattice by frustrating all 
the dual links intersecting a given surface $\Sigma$ bounded by $C$. 
These frustrated links modify the vacuum state so that the 
expectation value $\br P\kt_W$ of the plaquette, or better its spin 
counterpart described in Eq.~(\ref{wil}),  becomes a function 
of its relative position with respect to $W(C)$ and is related to 
the expectation value in the ordinary vacuum by
\eq
\br P\kt_W=\br W(C)\,P\kt/\br W(C)\kt~~.
\en
The difference between the expectation value of the plaquette in the 
vacuum modified by the presence of $W(C)$ and in the ordinary 
vacuum can be considered as a measure of the colour flux 
density of the flux tube. 
Choosing for instance as a probe a plaquette $P_\parallel$ 
parallel to the plane of the Wilson loop we can take 
\eq
\rho_\parallel(x,y,z)=\br  P_\parallel\kt_W-\br P\kt~~.
\label{density}
\en
Other orientations of the plaquette give approximately the same 
distribution (this is not the case for other  
systems \cite{dg,bss,tw}).  

It is important to stress that within our procedure we do not need to 
evaluate directly the correlation function $\br W(C) P\kt$ between the 
plaquette and the Wilson loop, which is very difficult in ordinary 
lattice gauge model simulations and must be handled with  special 
techniques like the cooling \cite{dg} or the smearing methods 
\cite{bss}. Actually we only have to evaluate, according to 
Eq.~(\ref{density}), 
the mean expectation value of the plaquette with two 
different actions: the ordinary action and the one modified by the 
Wilson loop. This allows us 
to estimate $\rho$ with high accuracy even with 
relatively modest computational means. A preliminary account of these 
numerical simulations has been reported in Ref.~\cite{cgmv}.

\vskip .2 cm
%%%%%%%%%%%%%%%%%%%%%%%%%%%%%%%%%%%%%%%%%%%%%%%%%%%%%%%%%
\centerline{\sl Tab.~II}
\begin{center}
\begin{tabular}{cccccc}
\hline
$\beta$&lattice size&$R$&$T$&\# data&$w^2_o$\\
\hline
0.7516&24\t24\t48&12&12&12500&19.0(1.1)\\
0.7460&22\t22\t40&11&11&35000&12.3(5)\\
0.7460&30\t30\t48&15&15&7000&15.6(8)\\
0.7460&60\t60\t48&30&30&5000&20.3(4)\\
0.7460&120\t120\t32&60&60&1000&24.8(6)\\
0.6543&64\t64\t32&32&32&7400&2.57(2)\\
0.6543&124\t124\t24&32&64&2000&2.79(5)\\
0.6543&96\t192\t32&32&96&1700&2.75(5)\\
0.6543&124\t124\t20&64&64&1000&2.95(3)\\
\hline
\end{tabular}
\end{center}
\vskip .2 cm
\begin{center}
{\it The first two columns show the $\beta$--values and the corresponding 
size of the lattices used in the present study, in the notation 
$L_x$~{\em x}~$L_y$~{\em x}~$L_z$ where $R$ and $T$ point in the x and y 
directions respectively. In the last four columns the size of the 
Wilson loops, the number of data and the corresponding estimates 
of the mean squared width of the flux tube are reported.}
\end{center}
%%%%%%%%%%%%%%%%%%%%%%%%%%%%%%%%%%%%%%%%%%%%%%%%%%%%%%%%%
\vskip .2 cm

We would also like to point out that the plaquette expectation values 
entering into Eq.~(\ref{density}), namely 
$\br  P_\parallel\kt_W$ (with the W--vacuum) and $\br P\kt$ 
(with the ordinary vacuum), are evaluated in distinct runs, so that
the corresponding measurements are completely uncorrelated.

We analyzed three distinct values of the coupling $\beta$  and 
different sizes and shapes of square and rectangular Wilson loops,
as can be seen from Tab.~II.

For each run with a W--vacuum, the Wilson loop is located in the $z_0$
plane. 
Cross--correlations in the measurements of the plaquette 
expectation values defined at different distances $|z-z_0|$ from the 
Wilson loop have to be carefully considered in this procedure.
To keep them under control we did five update sweeps of the 
lattice between two
measurements, and we measured only plaquettes in 
a fixed $z-z_0$ slice each time.
This means, for instance, that two measurements of the same observable 
were
separated by $5\times L_z$ sweeps, and there were 
at least $5$ sweeps between
measurements of different observables. From 
Tab.~II one can see that each run contained between 100 000 and 
240 000 sweeps.

Moreover, if one takes a plane parallel to the Wilson loop
and looks at the shape of $\rho$ as a function of the $x$--$y$ 
coordinates 
on this plane, one can see that $\rho$ is approximately constant, 
except in a region extending over a few lattice spacings from the 
border of the Wilson loop \cite{gliozzi}. Thus, in order to have a sufficiently
stable evaluation of the width at the center of the Wilson loop,
we  averaged the plaquette expectation values 
$\br  P_\parallel\kt_W$ over the $x$--$y$ directions, excluding a border
along the Wilson loop perimeter 
of size four lattice spacings across.

A typical shape of the distribution $\rho$ measured on the symmetry 
axis of the loop as a function of the distance $z-z_0$ from the loop 
plane is drawn in Fig.~2. 
The dotted straight line corresponds to the plaquette expectation 
value in the ordinary vacuum, $\br P\kt=0.8945(1)$ for $\beta=0.7460$ 
(while we obtained $\br P\kt=0.70852(2)$ at $\beta=0.6543$ and 
$\br P\kt=0.9108(2)$ at $\beta=0.7516$): it does not show appreciable 
finite size effects for the lattices we have considered.
The other two curves will be explained in the last section. \par

The mean squared width of the flux tube can be defined as
\eq
w^2_o=\frac{\int\,z^2\,\rho_\parallel\, dx\,dy\,dz}{\int\rho_\parallel\,  
dx\,dy\,dz}~~~.
\label{zw2}
\en
The integration region in the horizontal directions  is taken over 
the central plateau where $\rho$ is approximately constant as a function of 
$x$ and $y$.
The evaluation of the integral  was performed numerically and all the 
errors were evaluated with the binning procedure and the jackknife 
method.

It is important to stress that the numerical integrations are not 
reliable  if 
the asymptotic values of $\rho$ and $\br P\kt$ do not agree within 
errors.
To this end it is crucial to take large lattice sizes compared to the 
Wilson loops (see Tab.~II).

An  important feature of Eq.~(\ref{log}) is that it can be written 
in a universal, $\beta$--independent form by expressing all the 
dimensional quantities $w_o$, $R$ and $R_c$ in units of $\sqrt{\sigma}$. 
Accordingly, we report in Fig.~3  the  data of Tab.~II  
and also very accurate  data \cite{hp} for the mean squared width 
of fluid interfaces  in the  Ising model (black dots) for one of the 
couplings analyzed in our approach ($\beta=0.6543$ of the gauge model).
Note indeed that, according to Tab. I, the squared width of the flux tube 
with {\sl periodic} boundary conditions (which is the case of the fluid
interfaces) is almost the same of that with fixed b.c.

The straight line represents  a fit of all the data
to the two--parameter formula
\eq
\sigma w^2_o=a\,\log\,\sqrt{\sigma}R\,+\,b~~~.
\label{ab}
\en
We used the values $\sigma=0.0107(1) $ at $\beta=0.7516$ and 
$\sigma=0.0189(1)$ at $\beta=0.7460$ according to Ref. \cite{inter2}.
Far from the scaling region, the string  tension should be replaced by 
the stiffness $\kappa$. This is the case for $\beta=0.6543$ where we 
used $\kappa=0.234(9)$ according to Ref. \cite{hp}.
One gets $b=0.173\pm0.009~$ and $a=0.150\pm0.005$ in good agreement 
with the  theoretical value $a_{th}$ of the parameter $a$, fixed by 
Eq.~(\ref{log}) to be $a_{th}=\frac1{2\pi}\simeq0.159155~$.

Within high statistical accuracy these data clearly support a 
logarithmic widening of the flux tube in a range of quark separation $R$
over more than two orders of magnitude,  starting at about 
$\sqrt{\sigma}R_{min}\simeq 1.2$.

The physical adimensional quantity $\sqrt{\sigma} R_c$  is expressed in 
terms of $a$ and $b$ as $\sqrt{\sigma} R_c={\mathrm e}^{-b/a}$.
Using the fitted values of the parameters we get 
$\sqrt{\sigma} R_c=0.32\pm0.02$. 

In order to compare these results with analogous data for other gauge 
groups, we may, with an abuse of language, express $\sqrt{\sigma}$ in 
the physical units of the QCD string . Then the 
infrared logarithmic behaviour observed in Fig.~3 starts at least at
$R_{min}\simeq0.6$ fm while the maximal 
probed elongation of the flux tube corresponds to more than 100 fm. 
The data on $4D$ 
$SU(2)$ flux tubes \cite{dg,bss}  cover now a distance up to 2 fm, but 
are still affected by strong systematic errors and are compatible also 
with a constant width for distances larger than 1 fm. 

\section{Discussion}

The observed logarithmic widening of the flux 
tube exhibits, for the first time,  
a macroscopic effect due to the string--like motion of the colour flux 
tube. At present it is visible only in the $3D$ $\ZZ_2$ gauge system 
because only there it is possible to reach, using scaling and the 
duality property of this system, large separations of the quark sources. 
Nevertheless we expect a similar behaviour in all $3D$ and $4D$ 
gauge systems, because the appearance of string--like vibration modes of 
the colour flux tube is a universal property of the rough phase of 
whatever gauge model.

The agreement of the logarithmic growth of the mean squared width of 
the flux tube observed in $\ZZ_2$ gauge model with the predictions of 
the gaussian model seems  rather impressive: indeed the 
universal slope of the logarithmic growth has been checked in a range 
(more than two order of magnitude) which is unusually wide in 
numerical simulations. 

Admittedly, this prediction is not specific to the 
gaussian model: such a model 
is a member of a one--parameter family of conformal field theories, 
namely those with conformal anomaly $c=1$. These theories have the 
same local behaviour (hence produce the logarithmic growth of the mean 
squared width) but have different topological excitations described by 
winding modes around a circle of radius $r$. This is the parameter 
labelling the different members of the family and $r\to\infty$ 
corresponds to the gaussian limit. 

Also the agreement with some predictions which are more specific of the gaussian
model is rather good.
In particular, the scale $R_c$ of the logarithmic growth depends also on the 
shape $t=T/R$ of the rectangular Wilson loop. As a consequence, the difference 
$\Delta w^2(R)=w^2_{t}(R)-w^2_p(R)$ between  the mean squared width of 
the flux tube generated by the rectangular Wilson loop 
and that of the reference torus does not depend on $R$ but is a 
calculable function of the shape of $t$ as shown in Fig.~1. 
For fixed boundary conditions (dashed line) it reaches the asymptotic 
value of the infinite strip  for very large values of $t$.
The data extracted from numerical simulations in $\ZZ_2$ 
gauge model (black dots) seem to agree with this behaviour, even if the errors
are too large to reach a definite conclusion. The fact that the data seem a bit
lower than the expected values is probably due to a (non universal) finite 
size effect; indeed in the $2D$ $XY$ model with fixed b.c. such an effect can
be easily observed.

Another specific prediction of the gaussian model is that the scales $R_c$ for
fixed and periodic boundary conditions are almost the same. This fact is nicely
confirmed with high accuracy by Fig.3.

The only feature of the gaussian model which seems to slightly deviate form the
observed data concerns the distribution of the flux density as a function of 
the distance from the Wilson loop. It should be, according to 
Eq.~(\ref{gaussian}), a gaussian  $\rho_g(z)$. Although 
$\rho_g$ (dashed line in Fig.~2) reproduces the raw features of 
the data, the fit is not very good. A much better fit is obtained by
putting $\rho=\rho_g/(1-\gamma\rho_g)$, where  $\gamma$ is  another 
free  parameter (continuous line). 
This is suggested by the assumption that the flux tube does 
actually self--interact  and the interaction should be a power series 
in $\rho$.

In conclusion, the above  results seem to indicate that the free 
gaussian model is a good approximation of the string--like 
behaviour of the flux tube at least in the infrared limit. 
\newpage

{\bf Acknowledgements}
\vskip 0.3cm
We would like to thank M. Hasenbusch and K. Pinn for providing us with a 
very efficient non--local algorithm for the $2D$ $XY$ model. This work 
has been partly supported by the Ministero dell'Universit\`a e della 
Ricerca Scientifica.

\vskip .8 cm
%\newpage
\appendix{\Large {\bf{Appendix}}}
\label{hkscapp}
\vskip 0.5cm
\renewcommand{\theequation}{A.\arabic{equation}}
\setcounter{equation}{0}

The Green function of the $2D$ free massless theory on a rectangle of size 
$R\times T$ with periodic boundary conditions on both sides, as given in 
Eq.~(\ref{sol}), can be expanded as \cite{ahlfors}
\eq
\CG(z)=-\frac1{2\pi}\Re e\left\{\log z-\frac{(\Im m z)^2}{2RT}
-\sum_{k=1}^\infty G_k\frac{z^{2k}}{2k}\right\}
\en
with
\eq
G_k=\frac{2\zeta(2k)}{R^{2k}}E_{2k}(it)~~,~~t=T/R
\en
where $\zeta$ is the Riemann zeta function and the Eisenstein series 
have the following expansion
\eq
E_{2k}(\tau)=1+(-1)^k\frac{4k}{B_k}\sum_{n=1}^{\infty}\sigma_{2k-1}(n)
q^n~~~,~~~q={\mathrm e}^{2i\pi\tau}~~~.
\label{eis}
\en
$\sigma_k(n)$ denotes the sum of $k$th powers of the positive 
divisors of $n$ and $B_k$ are the Bernoulli numbers.

Since $\CG(z,z')$ is translationally invariant under  $z\to z+a$;  
$z'\to z'+a$, it is only a 
function of the difference $z-z'$; then the mean squared width as defined in
Eq.(\ref{w22}) $w^2$ is a constant with respect to $z$ that we denote with 
$w^2_{per}(R)$. Performing the integrations we get at once
\bea
2\pi\sigma w^2_{per}(R)=& \log\left({R\sqrt{1+t^2}}/{2\varepsilon}
\right)+\frac1{2t}\arctan t + \frac t2\arctan\frac1t-\frac32
-\frac{\pi t}{12}\nn
-&\sum_{k=1}^{\infty}\frac{(1+t^2)E_{2k}(it)}
{2kt(2k+2)!}\left((1+t^2)\pi^2\right)^kB_k\sin\left((2k+2)\arctan 
t\right)~~~.
\label{per}
\eea
This expansion converges only for $\sqrt{1+t^2}<2$ because there is a 
pole at $z=2i\pi$ in the complex plane of the variable 
$z=\pi(R+iT)/R$. However it is  possible to perform an analytic 
continuation which allows us to evaluate $w^2$ for any value of $t$,
 by resumming the potentially divergent part of the series.
Indeed exploiting the well--known expansion
\eq
\frac{z}{{\mathrm e}^z-1}=1-\frac{z}2-\sum_{k=1}^\infty(-1)^k
\frac{B_k}{2k!}z^{2k}~~~,
\label{bernoulli}
\en
we can easily derive the result
\eq
\sum_{k=1}^\infty(-1)^k
\frac{B_k}{(2k+2)!}\frac{z^{2k+2}}{2k}=\sum_{n=1}^\infty
\frac{{\mathrm e}^{-nz}-1}{n^3}+\frac{z^2}2\log z+\frac{\pi^2}6z
-\frac34z^2-\frac{z^3}{12}~~,
\label{resum}
\en
which is useful for large $\vert z\vert$. Its insertion in 
Eq.~(\ref{per}) gives the wanted resummation, once one realizes that 
$E_{2k}(it)=1+O({\mathrm e}^{-2\pi t})$. In particular, for large 
values of $t$ one gets the asymptotic expansion
\eq
2\pi\sigma w^2_{per}(R)\to
\frac\pi6t+\log\frac R{2\pi}~~,~~(t\to\infty)~~,
\en
which is the expected behaviour in the one--dimensional limit.

For the rectangle with fixed boundary conditions one can use the same 
procedure, but in this case, owing to the fixed boundary, the Green 
function $G_{R\times T}$, is no longer translationally invariant. 
For the $\varepsilon$--dependent part of 
the Green function it is advantageous to use 
the following representation 
\eq
G_{R\times T}(z,z')=\CG(z-z')+\CG(z+z')-\CG(z-\bar{z'})-
\CG(z+\bar{z'})~~~,
\en
where $\CG$ is the periodic Green function  for a torus of size 
$2R\times 2T$. Denoting with
$w^2_{fix}(R)$ the mean squared width evaluated at the symmetry point of the
Wilson loop we have
\bea
2\pi\sigma w^2_{fix}(R)=& \log\left(Rt/\sqrt{1+t^2}\varepsilon\right)\nn
-\sum_{k=2}^{\infty}&\frac{\pi^{2k}E_{2k}(it)}{2k(2k)!}B_k
\left[(1+t^2)^k\cos(2k\arctan t)-(-t^2)^{k}-1\right]\nn
&+\chi(t)-\sum_{k=2}^\infty\frac{3\pi^{k}E_{2k}(it)}
{kt(2k+2)!}B_kc_k\nn
-&\hskip-1 cm\frac{2\pi}{R^2T^2}\int_{R\times T}{\mathrm d}^2z
\int_{R\times T}{\mathrm d}^2z'\;G_{R\times T}(z,z')~~,
\label{fix}
\eea 
with
\bea
c_k=&3(3^{2k}+1)(1+t^2)^{k+1}\sin(2k+2)\arctan{t}-\nn
&(1+9t^2)^{k+1}\sin(2k+2)\arctan{3t}-(9+t^2)^{k+1}\sin(2k+2)\arctan{t/3}\\
\chi(t)=&\left[9f(R,T)-3f(3R,T)-3f(R,3T)+f(3R,3T)\right]/RT\nn
f(x,y)=&\frac{xy}4\log(x^2+y^2)+\frac{x^2}4\arctan\frac{y}x+
\frac{y^2}4\arctan\frac{x}y~~~.
\eea

Again the infinite sums can be resummed using Eq.~(\ref{bernoulli}).

The quadruple integral in Eq.~(\ref{fix}) has the following 
representation
\bea
\frac{2\pi}{R^2T^2}\int_{R\times T}{\mathrm d}^2z
\int_{R\times T}{\mathrm d}^2z'\;G_{R\times T}(z,z')=&
\frac\pi{6t}+\frac{31}{\pi^4t^2}\zeta(5)\nn
&-\frac{64}{\pi^4t^2}\sum_{n=1}^\infty\frac{(2n-1)^{-5}}{1+q^{2n-1}}~~~,
\eea
where $\zeta$ is the Riemann zeta function and $q=\exp(-\pi t)$.

A check of these formulas can be obtained by evaluating the 
$t\to\infty$ limit of Eq.~(\ref{fix}), giving
\eq
\lim_{t\to\infty}2\pi\sigma w^2_{fix}(R)=\log\left(2R/\pi\varepsilon\right)~~~,
\en
which coincides with the result for the infinite strip obtained 
by directly using the Green function of Tab.~I.

\vskip1.2cm
\begin{center}
{\bf  Figure Captions}
\end{center}
\vskip0.8cm
\begin{description}
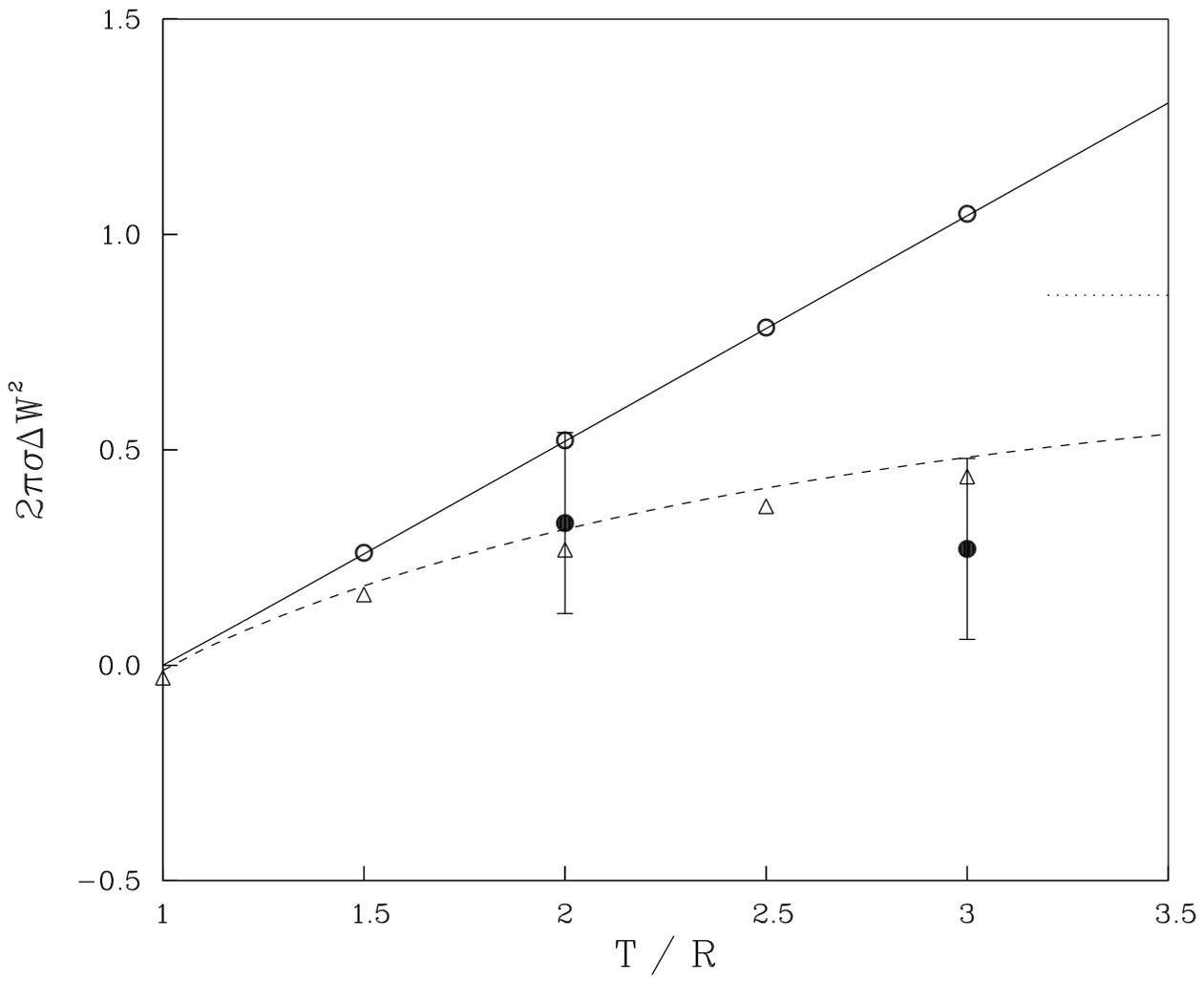
\item{Fig.~1.} Variation of the mean squared width of the flux tube 
as a function of the ratio $T/R$ calculated in the gaussian model 
and compared with numerical simulations in the
$2D$ $XY$ model (open circles and triangles for periodic and fixed 
boundary conditions respectively) and the $3D$ $\ZZ_2$ gauge theory
(black dots). The 
continuous and dashed lines correspond to periodic 
and fixed boundary conditions, respectively. The dotted line indicates the 
infinite strip limit.
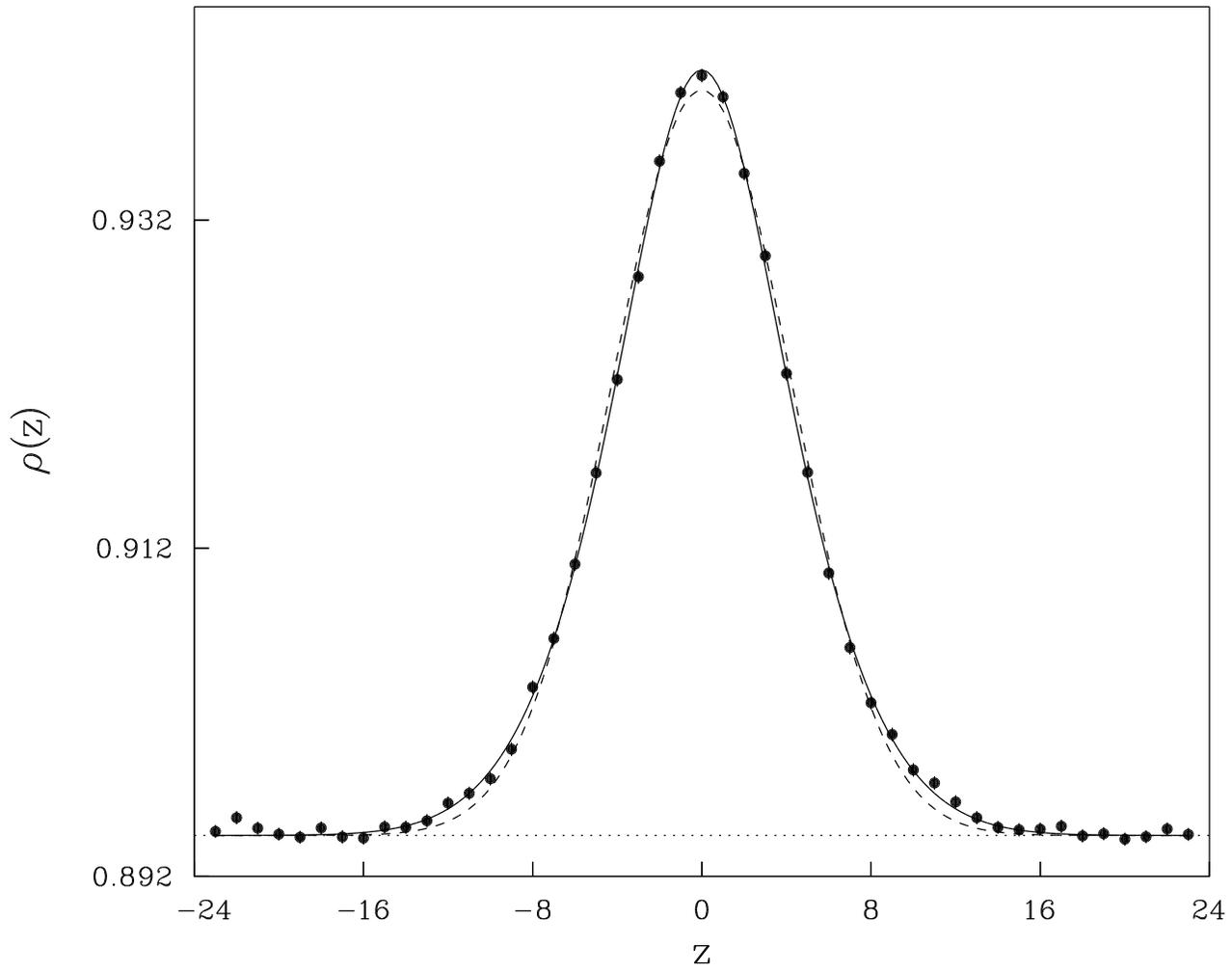
\item{Fig.~2.} Density of the flux tube generated by a $30\times30$ 
Wilson loop at $\beta=0.7460$. The dashed line is the fit to
the gaussian distribution of Eq.~(\ref{gaussian}). The continuous line 
is a fit to the function described in the text while the dotted line
is the plaquette expectation value in the ordinary vacuum.
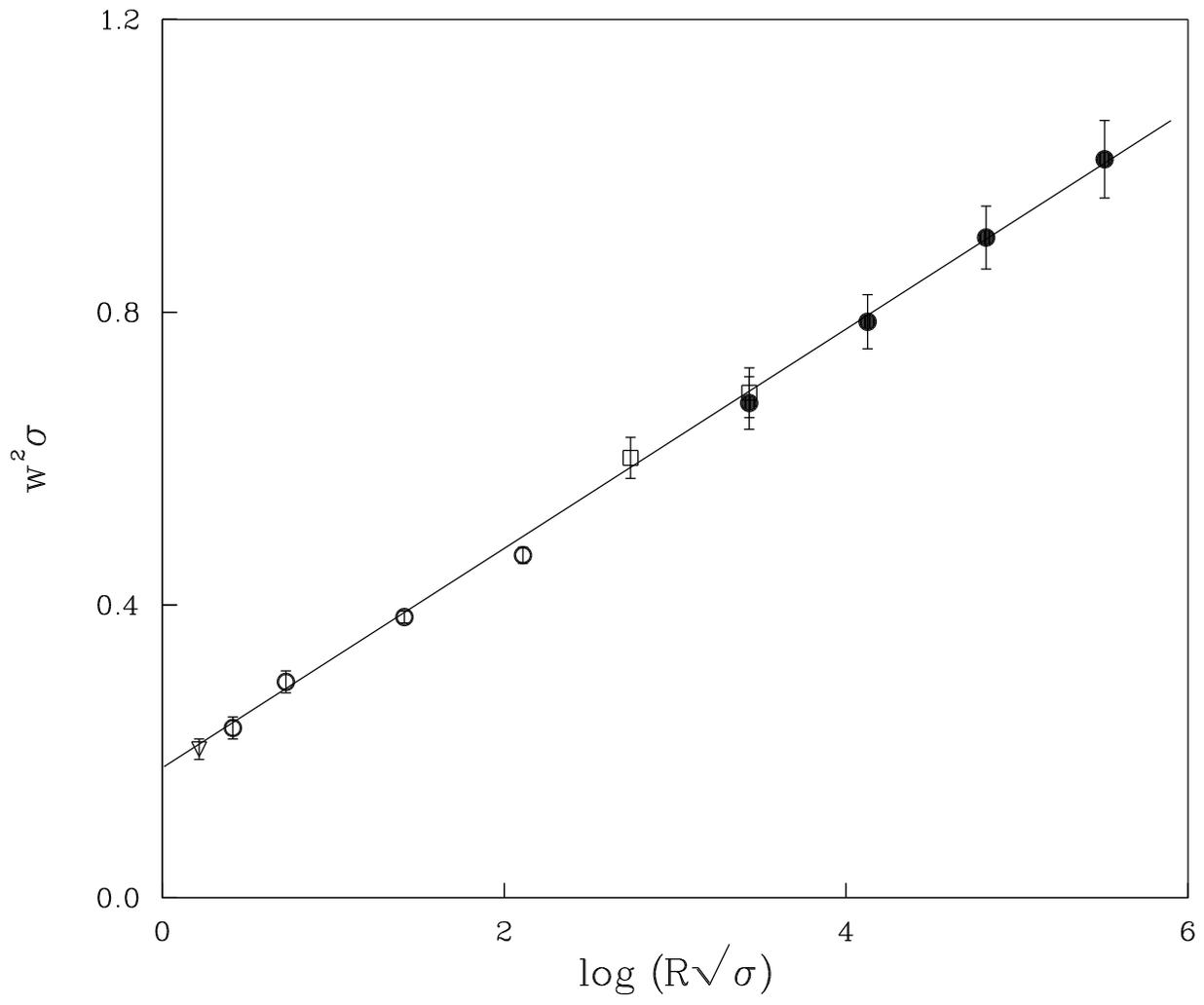
\item{Fig.~3.} Squared width of the flux tube in units of sigma. The open 
symbols are the data of the $\ZZ_2$ gauge theory and are taken from 
Tab.~II, while the black circles are data from the Ising interface taken 
from Ref.~\cite{hp}. 
\end{description}
\end{document}